\documentclass[12pt]{article}
\usepackage{epsfig,epsf}

\newcommand{\beq}{\begin{equation}}
\newcommand{\eeq}{\end{equation}}
\def\bea{\begin{eqnarray}}  \def\eea{\end{eqnarray}}

\def\be{\begin{equation}}
\def\ee{\end{equation}}
\def\bea{\begin{eqnarray}}
\def\eea{\end{eqnarray}}

\def\simge{\mathrel{%
   \rlap{\raise 0.511ex \hbox{$>$}}{\lower 0.511ex \hbox{$\sim$}}}}
\def\simle{\mathrel{
   \rlap{\raise 0.511ex \hbox{$<$}}{\lower 0.511ex \hbox{$\sim$}}}}
\def\del{\partial}

\begin{document}
\vspace*{2cm}
\begin{center}

{\Large \bf
From percolation to the Color Glass Condensate}\\[8mm]
{\bf  Elena Ferreiro}\par
{\it Departamento de
F\'{\i}sica de  Part\'{\i}culas, 
Universidad de Santiago de
Compostela,\\ 15782 Santiago de Compostela, Spain}


\vskip 1.0truecm
\end{center}
\begin{abstract}
I review the problem of parton saturation and its implications through three in principal 
different
approaches, but somewhat related: saturation in a geometrical approach, QCD saturation through the
Color Glass Condensate and perturbative Pomeron approach with initial conditions.
I also make some comments about how that could be related to a Quark Gluon Plasma formation.
\end{abstract}
\vskip 1.0truecm


\newpage

\section{Introduction}
In the recent experiments like DIS at HERA or the heavy-ion experiments at RHIC, and
also in expected LHC at CERN, the number of involved partons is very large,
due to the high energy and/or the high number of participants of those experiments.
These high parton densities should in principal lead
to an extremely huge multiparticle production, but experimentally we have seen that
this is not the case. So there should be a mechanism that reduces the number of created
particle. Two kind of phenomena have been proposed. On one side there is the possibility
of saturation in the initial state of the collision. On the othe side, there is the proposal
of the creation of a Quark Gluon Plasma,
and it has been presented as a final state interaction mechanism.
Here, I review the problem of parton saturation and its implications through three in principal
different
approaches, but somewhat related: saturation in a geometrical approach, QCD saturation through the
Color Glass Condensate and perturbative Pomeron approach with initial conditions.
I will also make some comments about how that could be related to a QGP formation.

\section{Geometrical approach to saturation}
\subsection{String models and percolation}

In many models of hadronic and nuclear collisions, color strings are exchanged between 
the projectile and the target. Those strings act as {\it color sources} of particles through the creation
of $q-{\bar q}$ pairs from the sea. 
The number of strings grows with the energy and with the number of nucleons of the 
participant nuclei. 


In impact parameter space these strings are seen as circles inside the total collision area.
When the density of strings becomes high the string color fields begin to overlap and 
eventually individual strings may fuse, 
forming a new string --{\it cluster}-- which has a higher color charge at its ends, 
corresponding to the summation of the color charges located at the ends of the original strings. 
The new string clusters break into hadrons according to their higher color. 
As a result, there is a reduction of the 
total multiplicity. 
Also, as the energy-momenta of the original strings are summed to obtain the energy-momentum of 
the resulting cluster, the mean transverse momentum of the particles created by those clusters
is increased compared to the one of the particles 
created from individual sources.

\begin{figure}
\centering\leavevmode
\epsfxsize=15cm\epsfysize=4cm\epsffile{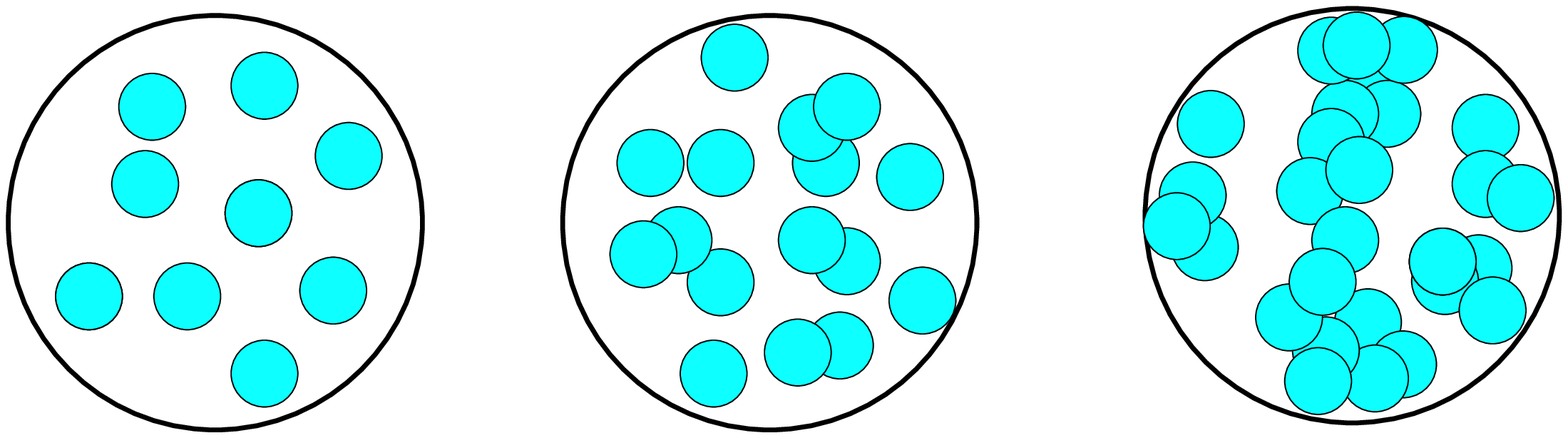}
\caption{From left to righ: 
Density of strings in the transverse space, from low energy and/or
low number of participants to high energies and/or high number of 
participants. In the last circle we show percolation.}
\label{fig1}
\end{figure}

As the number of strings increases, 
more strings overlap. 
Some years ago, it has been proposed in Ref. \cite{REF96} that
above a critical density of strings {\it percolation}
 occurs, so that paths of overlapping circles are formed through the whole collision area,
as it is represented in Fig. \ref{fig1}. 
Along these paths the medium behaves like a color conductor. 
Also in \cite{REF96}, we have made the remark that 
several fused strings can be considered as a domain of a {\it non thermalized Quark Gluon Plasma}. 
The percolation gives rise to the formation of a non thermalized 
Quark Gluon Plasma on a nuclear scale. 

Note that here we are not speaking about a final state interaction phenomenon, since there is no 
thermalization involved. In fact, what we are trying is
to determine under what conditions the initial state 
configurations can lead to color connection, and more specifically, 
if variations of the initial state can lead to a transition from 
disconnected to connected color clusters. 
The results of such a study of the pre-equilibrium state in nuclear collisions 
do not depend on the subsequent evolution and thus in particular not require any kind of 
thermalization.

The structural problem underlying the transition from disconnected to connected systems 
of many components is a very general one, ranging from clustering in spin systems to 
the formation of galaxies. 
The formalism is given by percolation theory, which describes geometric critical behavior.

\subsection{Percolation theory}

Consider placing $N$ small circular discs (color sources, strings or partons) 
of radius $r$ onto a large circular manifold (the transverse nuclear plane) of radius $R$; 
the small discs may overlap. 
With increasing density 
\be
\eta = \frac{N \pi r^2}{ \pi R^2}\  , 
\label{ec1}
\ee
this overlap will lead to 
more and larger connected clusters. 
The most interesting feature of this phenomenon is that the average cluster size 
increases very suddenly from very small to very large values. 
This suggests some kind of geometric critical behavior. In fact, 
the cluster size diverges at a critical threshold value $\eta_c$ of the density. This appearance
of an infinite cluster at $\eta= \eta_c$ is defined as percolation: the size of the cluster 
reaches the size of the system. 
$\eta_c$ has been computed using Monte Carlo simulation, direct connectedness expansion and 
other different methods. All the results are in the range  $\eta_c = 1.12 \div 1.175$.

In our model \cite{REF96} we had proposed a {\it fixed radius} for the independent 
color sources of 
$r=0.2 \div 0.25$ fm,
 that corresponds to a momentum around 1 GeV. 
This value has been obtained from 
Monte Carlo simulations in the framework of the String Fusion Model Code (SFMC) \cite{REFSFMC} 
made at SPS
energies. 
According to eq. (\ref{ec1}), in order to estimate the density $\eta$, one needs to know 
the number of sources $N$. In our model, it is obtained from the SFMC,
that, for nucleus-nucleus collisions, 
takes into account two contributions: one proportional to the number of participant
nucleons --valence-like contribution-- and another one proportional to the number of 
inelastic nucleon-nucleon collisions.
Note that $N$ will depend on the energy $\sqrt{s}$ (or equivalently, on $x$) and on the number of 
participant nucleons $A$, so in some way the condition to achieve percolation depends on 
$A$ and $s$, $\eta=\eta(A,x)$.
$\pi R^2$ corresponds simply to the nuclear overlap area, $S_A$, 
at the given impact parameter. This overlap area can be determined in 
a Glauber study, using Woods-Saxon nuclear profiles.
That leads to the following results: In our model,
at SPS energies, the critical threshold for percolation could eventually been 
achieved for the most central Pb-Pb collisions, and for sure in Au-Au central collisions at 
RHIC energies and even in p-p collisions at LHC energies.
Just as an example, in first approximation, one can estimates analytically that 
at very high energies, for central A-A collisions, if we take the number of initially created
sources as proportional to the number of collisions, 
$N \propto A^{4/3}$ and the nuclear overlap area $S_A= \pi R^2 \propto A^{2/3}$ 
then 
\be
\eta=\frac{N \pi r^2}{ \pi R^2}=
\frac{N_{sources}
S_1}{S_A} \propto A^{2/3}\ .
\label{ec2}
\ee

\subsection{%
The size of the color sources}

Let us now introduce a new question:
If you look at a fast nucleon coming at you, what do you see? 
It depends on who is looking. Another nucleon sees a disc of radius $r \simeq 1$ 
fm and a certain greyness. 
A hard photon, with a resolution scale $Q^{-1} << 1$ fm, 
sees a swarm of partons. 
How many there are depends on the resolution scale: 
given a finer scale, you can see smaller partons, and there are more the harder you look.
The partons in a nucleon have a transverse size $r_T$ 
determined by their transverse momentum $k_T$ , with $r_T \sim 1/k_T$ . The scale $Q^{-1}$ 
specifies the minimum $k^{-1}_T$ resolved, so the probing photon sees all partons in the range 
$0 \le k_T \le Q$, or equivalently it sees all the partons with a radius $r_T \ge Q^{-1}$.

So the partonic size is through the uncertainty relation determined by its 
average transverse momentum, 
$r^ 2 \sim 1/<k_T^2>$, for a given resolution scale, $<k_T^2> \sim Q^2$.

\begin{figure}
\centering\leavevmode
\epsfxsize=15cm\epsfysize=4cm\epsffile{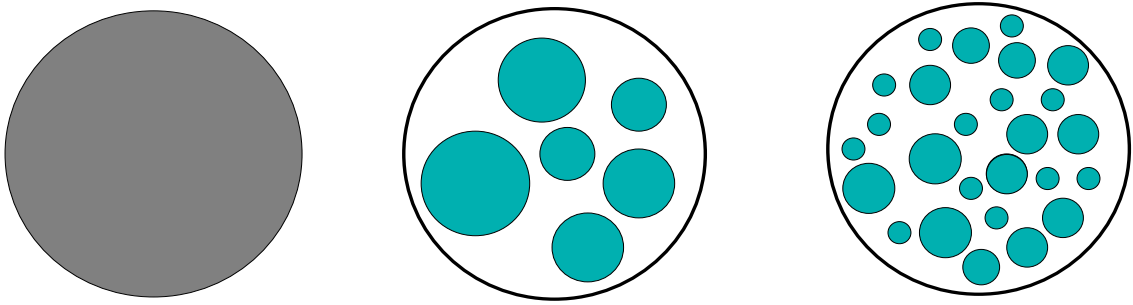}
\caption{From left to righ:
The structure of an incoming nucleon seen for increasing resolution.}
\label{fig2}
\end{figure}

\begin{figure}
\centering\leavevmode
\epsfxsize=15cm\epsfysize=15cm\epsffile{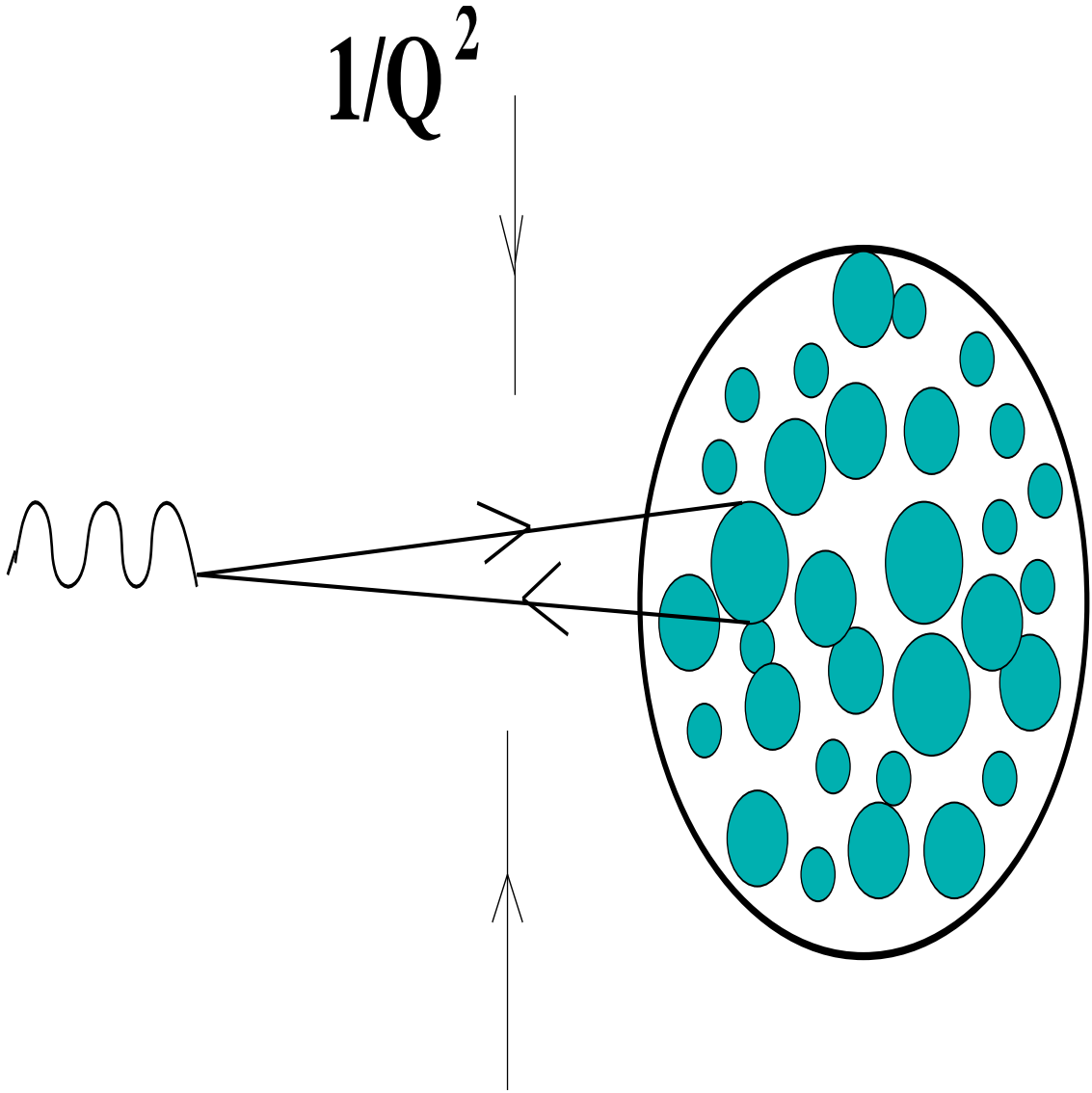}
\caption{
The resolution scale.}
\label{fig3}
\end{figure}

These ideas are illustrated in Figs. \ref{fig2} and \ref{fig3}.

In order to know if the percolation density is achieved in a collision,
we need to compute our eq. (\ref{ec1}). Then it is necessary to know 
the number of initially created sources (partons or strings, it depends on the
kind of model we are using) $N$, and the size of those sources $r$. Remember that in the previous
section we have used a fixed size of $r=0.2 \div 0.25$ fm that corresponds to a moment of 1 GeV.
In \cite{REF96} $N$ is calculated in the framework of the SFMC.

In \cite{REFSATZ}, the number of sources $N$ is calculated through the Wonded Nucleon Model for 
nucleus-nucleus collisions, using P.D.F.'s
for the nucleonic density.
The authors use here an slightly different strategy: Instead of fixing the radius of the initially
created sources, they estimate the momentum of those sources that will lead to percolation.
They transform then eq. (\ref{ec1}) into the following one:
\be
\eta_c = \frac{N \pi r^2_c}{\pi R^2}= \frac{N}{Q_c^2 R^2}\ .  
\label{ec3}
\ee
The condition for percolation, taking $\eta_c=1.12$
will be then:
\be
Q_c^2=\frac{N}{1.12 R^2}\ .
\label{ec4}
\ee
They find that for central Pb-Pb collisions at SPS energies, 
the critical momentum for percolation is $Q^2_c \approx 1$ GeV$^2$, in accordance with our result.
For central Au-Au collisions at RHIC, $Q^2_c \approx 2.5$ GeV$^2$.
Note that again the condition to achieve percolation depends on $A$ and $\sqrt{s}$,
$Q_c=Q_c(A,x)$. 

Beyond the percolation point, one has a condensate, containing 
interacting and hence color-connected sources of all scales $k_T \le Q$. 
The percolation point thus specifies the onset of color deconfinement; 
it says nothing about any subsequent thermalization.

\section{The Color Glass Condensate}

Now we arrive to another approach, the QCD saturation through the formation 
of a Color Glass Condensate (CGC) \cite{REFCGC}. The idea is the following:
At high energy, the QCD cross-sections are controlled by small longitudinal
momentum gluons in the hadron wave function, whose density grows rapidly with
increasing energy or decreasing $x$, due to the enhancement of radiative
process. If one applies perturbation theory to this regime, one finds
that, by resumming dominant radiative corrections at high energy, the BFKL
equation leads to a gluon density that grows like a power of $s$ and in
consequence
to a cross-section that violates the Froissart bound.
Nevertheless,
 the use of perturbation theory to high-energy problems is not obvious. In
fact, the BFKL and DGLAP equations are linear equations that neglet the
interaction among the gluons. With increasing energy, recombination effects
--that are non-linear--
favored by the high density of gluons should become more important and lead
to an eventual {\it saturation} of parton densities.

These effects become important
when the interaction probability for the gluons becomes of order one.
Taking $\frac{\alpha_s N_c}{Q^2}$ as the transverse size of the gluon and
$\frac{x G(x,Q^2)}{\pi R^2}$ as the density of gluons, the interaction
probability is expressed by
\be
\frac{\alpha_s N_c}{Q^2}\,\,\times\,\,
\frac{x G(x,Q^2)}{\pi R^2}\ .
\label{ec5}
\ee
Equivalently, for a given energy, saturation
occurs for those gluons having a sufficiently large transverse size $r_\perp^2
\sim 1/Q^2$, larger than some critical value $1/Q_s(x,A)$. So the phenomenon
of saturation introduces a characteristic momentum scale,
the {\it saturation momentum} $Q_s(x,A)$, which is a measure of the
density of the saturated gluons, and grows rapidly with $1/x$ and
$A$ (the atomic number).
The probability of interaction
--that can be understood as "overlapping" of the gluons in the transverse space--
becomes of order one for those gluons with
momenta $Q^2 \simle Q_s(x,A)$ where
\be
Q^2_s(x,A)=\alpha_s N_c \ \frac{x G(x,Q^2_s)}{\pi R^2} \equiv
\frac{({\rm color\,\, charge})^2}{{\rm area}}\ .
\label{ec6}
\ee

For $Q^2\simle Q^2_s(x,A)$, the non-linear effects
are essential, since they
 are expected to soften the growth of the gluon distribution
with $\tau\equiv \ln(1/x)$.
For a nucleus, $x G_A(x,Q^2_s)\propto A$ and
 $\pi R^2_A\propto A^{2/3}$, so eq.~(\ref{ec6})
predicts
$Q^2_s\propto A^{1/3}$. One can estimate the saturation scale by
inserting the BFKL approximation into
eq.~(\ref{ec6}). This gives
(with $\delta\approx 1/3$ and $\lambda\approx c\bar\alpha_s$ in a
first
approximation):
\be
\label{ec7}
Q^2_s(x,A)\,\,\sim\,\,A^{\delta}\, x^{-\lambda}\,,
\ee
which indicates that an efficient way to create a high-density
environment is to combine large nuclei with moderately small values of $x$,
as it is done at RHIC. In fact the estimated momentum for saturation at RHIC
will be $Q_s= 1 \div 2$ GeV, in accordance with the result of the previous section.

This equation also shows that for sufficiently large energy or $x$ small
enough, $Q^2_s(x,A) \gg \Lambda^2_{QCD}$ and $\alpha_s(Q_s) \ll 1$, which
characterizes
the regime of weak coupling QCD. But although the coupling
is small, the effects of the interactions are amplified by the large gluon
density: at saturation, $ G_A(x,Q^2_s) \sim 1/\alpha_s(Q_s) \gg 1$, so
the gluon modes
have large occupation numbers, of  order $1/\alpha_s$ (corresponding
to strong classical fields $A\sim 1/g$),
which suggests the use of semi-classical methods.

Note that in the string models described above, the formation of new strings --sources-- through 
the creation of $q-{\bar q}$ pairs from the sea is a non-perturbative way of gluon emission.

One can write a classical effective theory based on this general idea: the
fast partons --valence quarks with large longitudinal momentum-- are considered
as a {\it classical source} that emits soft gluons
 --with smaller longitudinal
momenta-- which are treated as {\it classical color fields} ${A[\rho]}$.
The fast partons move nearly at the speed of light in the positive $x^+$
direction, and generate a color current
$J^\mu=\delta^{\mu +}\!\rho$. By Lorentz contraction, the support
of the charge density $\rho$ is concentrated near the light-cone longitudinal
coordinate $x^-=0$. By Lorentz time
dilation, $\rho$ is independent of the
light-cone time
$x^+$.

The Yang Mills equation describing the soft gluon dynamics reads
\be
D_\nu F^{\nu \mu}\,\,=\,\,\delta^{\mu +}\rho(x^-,{\bf x})\ .
\label{ec5cgc}
\ee
Physical quantities, as the unintegrated gluon distribution, are obtained as
an average over $\rho$:
\be
< A^i(X) \, A^i(Y) >_x\,=\int D[\rho]\,\,W_x[\rho]\,\,
{A}^i[\rho](X)\,{A}^i[\rho](Y)\ ,
\label{ec6qgc}
\ee
where $A^i(X)$ corresponds to the classical solution for a given $\rho$,
and $W_x[\rho]$ is a gauge-invariant weight function for $\rho$.
What we are doing is a kind of
Born-Oppenhaimer approximation:
first, we study the dynamics of the classical
fields (Weizsacher-William fields)
for a given configuration $\rho$ of the color charges,
and second,
we average over all possible configurations.

For the classical solution we find:
\be
F^{+i}(x^-,x_\perp)= \delta(x^-) \frac{i}{g} V(x_\perp) (\partial^i
V(x_\perp)^\dagger)=\partial^+ A^i\ ,
\label{ec7acgc}
\ee
where $V(x_\perp)$ is the Wilson line
\be
V^\dagger(x_\perp)\equiv{\rm P} \!\exp
 \Big\{
ig\! \int\! dx^- \!A^+(x^-,{\bf x})
 \Big\}
\label{ec7cgc}
\ee
and $A^+[\rho]$ is the solution of the equation of motion (\ref{ec5cgc}) in the
covariant gauge: $- \nabla^2_\perp A^+=\rho$.

The weight function $W_\tau[\rho]$ is obtained by integrating out the
fast
partons, so it depends upon the rapidity scale $\tau={\rm ln}(1/x)$ at which
one considers the effective theory.
This can be taken into account via a one-loop background field calculation, and
leads to a renormalization group equation (REG) for $W_\tau[\rho]$
which shows how the correlations of $\rho$ change with increasing $\tau$.
Schematically:
\be
{\del W_\tau[\rho] \over {\del \tau}}\,=\,
 {1 \over 2} \int_{x_\perp,y_\perp} \,{\delta \over {\delta
\rho_x^a}}\,\,\chi_{xy}^{ab}[\rho]\,\,{\delta \over {\delta\rho_y^b}}\,\,
W_\tau[\rho]\ .
\label{ec8cgc}
\ee
This is a functional diffusion equation, where the kernel
$\chi[\rho]$ plays the role of the
diffusion coefficient in the functional space spanned by
$\rho(x^-,x_\perp)$.
This equation encompasses previous evolution equations \cite{REFBKW}
developed by Balitsky,
Korchegov and Weigert.
This kernel is positive definite and non-linear in $\rho$
to all orders. It depends upon $\rho$ via the Wilson line (\ref{ec7cgc}).

As it has been said above, the physical quantities, as the gluon  density, are
obtained as an
average over $\rho$:
\be
n(x, k_\perp)\,\equiv\,\frac{1}{\pi R^2}\,
\frac{{\rm d}N}{{\rm d}{\tau}\,{\rm d}^2 k_\perp}\,\propto\,
\langle F^{+i}(k_\perp) F^{+i}(-k_\perp)\rangle_{x}\ .
\label{ec9cgc}
\ee
In order to calculate this average, we need to solve the
REG (\ref{ec8cgc}).
Approximate solutions to this equation can be obtained in two limiting cases:

$\bullet$ At low energy, or large transverse momenta $k_\perp^2 \gg Q_s^2(x)$,
we are in a dilute regime where fields and sources are weak, and
the Wilson lines can be expanded to lowest order in $A^+$,
$V^\dagger(x_{\perp})\approx 1 + ig \int dx^- A^+ (x^-,x_{\perp}) $.
In this case, the REG equation reduces to the BFKL equation, and the gluon
density of eq. (\ref{ec9cgc}) grows both with $1/k_\perp^2$
and $1/x$ (Bremsstrahlung):
$n(x,k_\perp)\,\sim\,\frac{1}{k_\perp^2}\,\frac{1}{x^{\omega\alpha_s}}$.

$\bullet$ At high energies, or low momenta $k_\perp^2 \simle Q_s^2(x)$,
the color fields are strong, $A^+ \sim 1/g$, so the
Wilson lines rapidly oscillate and
average away to zero: $V\approx V^\dagger \approx 0$. Then the
kernel $\chi$ becomes independent of $\rho$, and we obtained a gluon density
that increases linearly with the evolution "time" $\tau=ln(1/x)$:
$n(x,k_\perp)\,\sim\,{1\over \alpha_s}\,
\ln{Q_s^2(x)\over k_\perp^2}\, \propto \,\,\ln{1\over x}$. 
That is, we find {\it saturation} for the gluon density, that grows
logaritmically with the energy since $\tau \sim \ln s$: unitarity is
restored.

We call the high density gluonic matter at small-$x$ described by this
effective theory a {\it Color Glass Condensate}:
{\it Color} since gluons carry color under $SU(N_c)$;
{\it Glass} since
we have classical coherent fields which are frozen over the typical time scales for 
high-energy scattering, 
but randomly changing over larger time scales. 
So we have a random distribution of time-independent
color charges
which is averaged over in the calculation
of physical observables, in order to have a gauge independent formulation
--in analogy to spin glasses--;
and {\it Condensate} because at saturation the gluon density is of
order $1/\alpha_s$, typical of condensates, so we have a system of saturate
gluons that is a Bose condensate.

\section{Perturbative QCD pomeron with saturation in the initial conditions}

\begin{figure}[ht]
\unitlength=1mm
\special{em:linewidth 0.4pt}
\linethickness{0.4pt}
\begin{picture}(93.00,129.33)
\put(30.33,129.33){\line(4,-5){30.00}}
\put(91.67,129.33){\line(-5,-6){31.00}}
\put(60.33,127.33){\line(-1,-1){16.00}}
\put(60.33,91.33){\line(0,-1){24.00}}
\put(60.33,67.67){\line(-5,-6){28.67}}
\put(60.33,67.33){\line(1,-1){32.67}}
\put(76.00,51.33){\line(-2,-3){11.67}}
\put(45.33,80.33){\line(1,0){30.33}}
\end{picture}
\caption{A typical diagram for the inclusive cross-section in nucleus-nucleus
collisions.}
\label {fig4}
\end{figure}
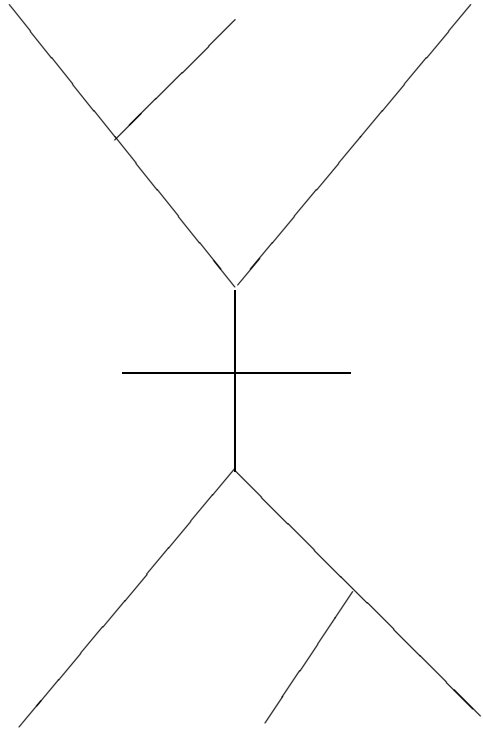

Let us now try a different approach \cite{REFMIJAILPAJ}. 
Consider now the nucleus-nucleus interaction as
governed by the exchange of pomerons, as it is represented in Fig. \ref{fig4}.
Its propagation is  governed
by the BFKL equation \cite{REFBFKL}. Its interaction is
realized by the triple
pomeron vertex. Equations
which describe nucleus-nucleus interaction in the perturbative QCD
framework have been obtained in ~\cite{REFBRA2}.
They are quite complicated and
difficult to solve, but
they will be
not needed for our purpose. Knowing that the AGK rules are satisfied
for the diagrams with BFKL pomerons interacting via the triple
pomeron vertex ~\cite{REFBARTELS} and using arguments of ~\cite{REFCIA} it is
easy to conclude that
 the inclusive cross-section will 
be given by the convolution of two sums of fan diagrams propagating
from the emitted particle towards the two nuclei. 

Taking $A=B$ and constant nuclear density for $|b|<R_A$, one can find
the inclusive cross-section in perturbative QCD as 
\be
I_{A}(y,k)=A^{2/3}\pi R_0^2\frac{8N_c\alpha_s}{k^2}\int
d^2re^{ikr}[\Delta\Phi_A(Y-y,r)]
[\Delta\Phi_A(y,r)],
\label{ec1p}
\ee
where $\Delta$ is the two-dimensional Laplacian and $\Phi(y,r)$ is the
sum
of all fan diagrams connecting the
pomeron at rapidity $y$ and of the transverse dimension $r$ with the
colliding nuclei, one at rest and the other at rapidity $Y$.
The function $\phi_A(y,r)=\Phi(y,r)/(2\pi r^2)$, in the momentum space,
 satisfies the well-known
non-linear BK equation ~\cite{REFBKW,REFBRA1}
\be
\frac{\partial\phi(y,q)}{\partial \bar{y}}=-H\phi(y,q)-\phi^2 (y,q),
\ee
where $\bar{y}=\bar{\alpha}y$, $\bar{\alpha}=\alpha_sN_c/\pi$,
$\alpha_s$ and $N_c$ are the strong coupling constant and the number
of colors,
respectively, and $H$ is the BFKL Hamiltonian. This equation has to be solved
with the initial condition at $y=0$ determined by the color dipole
distribution in the nucleon smeared by the profile function of the
nucleus.

We can take the initial condition in accordance with
the Golec-Biernat distribution \cite{REFGBW}, that takes into account
saturation according to the CGC, duly generalized for the nucleus:
\be
\phi(0,q)=-\frac{1}{2}a\,{\rm Ei}
\left(-\frac{q^2}{0.3567\, {\rm GeV}^2}\right),
\ee
with
\be
a=A^{1/3}\frac{20.8\, {\rm mb}}{\pi R_0^2}.
\ee

Then we can find the following result:
One can observe that whereas at relatively small momenta the inclusive
cross-sections are proportional to $A$, that is to {\it the number of
participants}, at
larger momenta they grow with $A$ faster, however  noticeably slowlier
than the number of collisions, approximately as $A^{1.1}$.
The interval of momenta for which $I_A\propto A$ is growing with energy,
so that one may conjecture that at infinite energies all the spectrum
will be proportional to $A$.

One can also compare \cite{REFMIJAIL} the results obtained on the basis of the expression for the 
inclusive cross-section which follows from the AGK rules applied to the diagrams 
with QCD pomerons interacting via the three-pomeron coupling \cite{REFBRA2}
with a somewhat different expression obtained from the color dipole picture \cite{REFKT}. 
These two formally different expressions lead to  identical results at momenta 
of the order or higher than the value of the so-called saturation momentum $Q_s$. 
At momenta substantially lower than $Q_s$ 
the color dipole cross-sections differ from the ones from the AGK rules 
by a universal constant factor $< 0.8 \div 0.9$. 
In both cases the spectra at momenta below $Q_s$ 
are found to be proportional to the number of participants ($A$ for collisions of identical nuclei) 
and not to the number of collisions $A^{4/3}$. 
Since $Q_s$ grows with energy very fast, the region where the spectra are 
proportional to  $A$ extends with energy to include all momenta of interest. 
At momenta greater than $Q_s$ the spectra grow faster than $A$ but still much slowlier than $A^{4/3}$
 (a numerical fit gives something like $A^{1.1}$).

\begin{figure}
\centering\leavevmode
\epsfxsize=15cm\epsfysize=15cm\epsffile{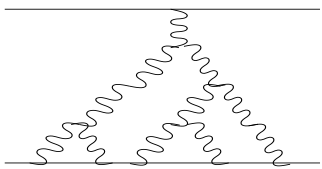}
\caption{
Fan diagram with triple pomeron interaction.}
\label{fig5}
\end{figure}

Note that in the string models, like Quark Gluon String Model or Dual Parton Model \cite{REFDPM}
each string corresponds to the exchange of pomerons, and the interaction among the strings 
would correspond to interaction among the pomerons through the triple pomeron vertex.
In particular, in the DPM some kind of saturation is also included through the {\it shadowing} in the 
initial state of the collision \cite{REFSHADOW}. In fact, in absence of shadowing,
the DPM at high energies leads
 to multiplicities that scale with 
the number of binary collisions rather than to a scaling with the number of participants. This is a general property of 
Gribov's Reggeon Field Theory which is known as AGK cancellation, 
analogous to the factorization theorem in perturbative QCD and valid for 
soft collisions in the absence of triple Pomeron diagrams.
Because of this, a dynamical, non-linear shadowing has been included in the DPM. It is determined
in terms of 
the diffractive cross-section. It is controlled by triple pomeron diagrams, shown 
in Fig. \ref{fig5}, and it should lead to 
saturation as $s \rightarrow \infty$. In fact, the shadowing corrections in this approach have a positive
contribution to diffraction and a negative one to the total cross-section.

\section{Conclusions}

We have compared different models that takes into account saturation in different ways:
from the semi-phenomenological
fusing color sources picture for the soft domain including percolation, 
the QCD saturation through the Color Glass Condensate
and those which
follow from the pomeron approach, perturbatively
derived from QCD, taking into account saturation in the initial conditions.

In fact, it seems that the exchanged of
elemental objects, color sources --strings, partons or pomerons--, should lead to a saturation
in the initial conditions when the densities are high enough.

I would like to finish making some remarks. Concerning
 the relation between percolation and thermal phase transitions, 
some thermal critical behavior, such as the magnetization transition for ferromagnetic spin systems, 
can be equivalently formulated as percolation. 
However, percolation seems to be a more general phenomenon and in particular can occur even 
 when there is no thermal critical behavior. 
A specific example of this is the 
Ising model in a non-vanishing external field, 
which has a percolation transition even though there is no magnetization transition.

In nuclear collisions there is indeed, as function of parton or string density, 
a sudden onset of large-scale color connection. 
There is a critical density at which the elemental objects 
form one large cluster, losing their independent existence and 
their relation to the parent nucleons. 
Percolation would correspond to the onset of 
color deconfinement and although it may be a prerequisite 
for any subsequent QGP formation, it does not require or imply any kind of thermalization.

The ancient idea about the formation of an ideal QGP --weakly coupled plasma of 
quarks and gluons-- above a critical 
temperature $T_c \sim 160$ MeV has changed: It seems that for moderate temperatures $(1 \div 3)\, T_c$,
accessible at RHIC, the plasma is predicted by non-perturbative lattice to be strongly coupled (sQCD),
and that its source could be a saturated initial state \cite{REFSQGP} of the type I have described here.

\section*{Acknowledgments}

I would like to thank C. Pajares and the group of 
Laboratoire Leprince-Ringuet, Ecole Polytechnique, CNRS-IN2P3 of France
for useful and fruitful discussions.


\newpage

\ \ \

\vskip 3.0truecm

\ \ \ 

\newpage

\ \ \

\ \ \ 

\newpage

\ \ \

\newpage

\newpage

\ \ \

\end{document}